\documentclass[aps,prl,reprint,superscriptaddress,amssymb,floatfix,showpacs]{revtex4-1}
\usepackage{amsmath}
\usepackage{amsfonts}
\usepackage{verbatim}
\usepackage{amsthm}
\usepackage{amssymb}
\usepackage{graphicx}
\usepackage{array}
\usepackage{enumitem}
\usepackage{pdfpages}
\usepackage{pgffor}
\newcolumntype{P}[1]{>{\centering\arraybackslash}p{#1}}
\newcolumntype{M}[1]{>{\centering\arraybackslash}m{#1}}

\makeatletter
\AtBeginDocument{\let\LS@rot\@undefined}
\makeatother

\newcommand{\fig}{Fig.\ }

\newcommand{\eqn}{Eq.\ }
\newcommand{\eqns}{Eqs.\ }

\newcommand{\kmax}{k_{\text{max}}}
\newcommand{\NEO}{\bar{N}}

\begin{document}

\title{Family-size variability grows with collapse rate in Birth-Death-Catastrophe model}

\author{N. Dori}
\affiliation{Gonda Brain Research Center, Bar-Ilan University, Ramat Gan, Israel}

\author{H. Behar}
\affiliation{Department of Biology, Stanford University
Stanford, CA 94305-5020}

\author{H. Brot}
\affiliation{Boston Children’s Hospital, Harvard Medical School, 3 Blackfan Circle, Boston, MA
Center for Polymer Studies and Physics Department, Boston University, Boston, MA, 02215, USA}

\author{Y. Louzoun}
\email[Corresponding author: ]{louzouy@math.biu.ac.il}
\affiliation{Gonda Brain Research Center, Bar-Ilan University, Ramat Gan, Israel}
\affiliation{Department of Mathematics, Bar-Ilan University, Ramat Gan, Israel}

\date{\today}


\MessageBreak

\begin{abstract}
Forest-fire and avalanche models support the notion that frequent catastrophes prevent the growth of very large populations and as such prevent rare large-scale catastrophes. We show that this notion is not universal. A new model class leads to a paradigm shift in the influence of  catastrophes on the family-size distribution of sub-populations. We study a simple population dynamics model where individuals, as well as a whole family, may die with a constant probability, accompanied by a logistic population growth model. We compute the characteristics of the family-size distribution in steady-state and the phase diagram of the steady-state distribution, and show that the family and catastrophe size variances increase with the catastrophe frequency, which is the opposite of common intuition. Frequent catastrophes are balanced by a larger net-growth rate in surviving families, leading to the exponential growth these families. When the catastrophe rate is further increased, a second phase transition to extinction occurs, when the rate of new families creations is lower than their destruction rate by catastrophes.
\end{abstract}


\maketitle


Catastrophes leading to partial or total population extinction are common in nature. From forest fires to collapsed markets, catastrophes have a crucial effect on the population dynamics of human beings. Accordingly, catastrophes have been studied extensively  in multiple contexts, including  ecology \cite{weng2012ecosystem, leite2012stationary, mawby1989endemic, turcotte1999self, chen2014model, wilcox2003effect, petraitis2013multiple, scheffer2001catastrophic, ives2004food, hart2014reconstructing, tuljapurkar2013population} and economics \cite{froot2008pricing, he2012model, enjolras2012combining,  nowak2013pricing, falco2014crop}.

Theoretic models of catastrophes have focused on self-organized criticality (SOC) models \cite{pruessner2012self}, such as the forest-fire  \cite{malamud1998forest, drossel1992self} and sand-pile models \cite{bak1988self,kadanoff1989scaling}. In these models, steady-state size distributions are famously characterized by an inverse relation between catastrophe frequency and catastrophe size.

This inverse relation is intuitive, and may be easily explained, as was done in the context of a simple spatial model of forest-fires \cite{malamud1998forest, turcotte1999self}, as well as more complex parallel models \cite{yoder2011forest, clar1994scaling, clar1996forest,noel2013controlling}.  In the simplest forest-fire model, trees are randomly planted on a grid at a constant rate, and sparks that can induce forest-fires are randomly ignited. The probability of a forest-fire scales like a power-law of the fire area. At low spark frequencies the number of forest-fires is small, but the burnt area is large, since the clusters of planted trees can percolate and spread over large areas. At high spark frequencies, the opposite occurs. Thus, in these models, more catastrophes are associated with a lower trees cluster-size, and catastrophes of smaller size. The strategy of allowing small forest-fires in order to prevent large ones is an accepted approach to fire prevention \cite{noel2013controlling}.

The well-established inverse relation between frequency and severity of catastrophes in SOC models may lead to the belief that catastrophes prevent the growth of very large family-sizes, or alternatively, major market crashes in economics. However, in both population dynamics and economics, catastrophes are very frequent and family-size distributions have a fat-tail \cite{stanley2002quantifying, fu2005growth, huynen1998frequency, koonin2002structure, qian2001protein, luscombe2002dominance}. Thus, frequent catastrophes (e.g. population extinction, or a collapse of a large company) do not seem to prevent the growth of fat-tailed family size distributions. Moreover, there is currently no good theory for the effect of such catastrophes in non-spatial models or models that do not have a limited local capacity.

In such models, the intuitive inverse relation between catastrophe frequency and the frequency of large families may fail.  Indeed, we here show that a new class of systems emerges from population dynamics of non-spatial models with catastrophes, in which higher catastrophe rates are correlated with a more inhomogeneous family-size distribution and more severe crashes. We study a simple,  solvable,  birth-death-innovation process that exhibits a direct relation between catastrophe frequency and catastrophe severity. While the model contains a minor modification of the reactions, the dynamics are governed by completely different phase transitions than the regular BDIM. We further show that applying the novel principles of the current model to other systems leads to novel dynamics in different systems, including network dynamics and spatial birth-death models.

 The current model has a limited total capacity that can be set to be arbitrarily large. However, beyond that it has no limit on the size of each family. New families are created by "mutations", and catastrophes are the annihilation of an  existing family.  In order to estimate the effect of catastrophes on the population structure, we analyze the family-size distribution. Such distributions have been studied mainly in duplication, loss and change (DLC) models or birth, death, innovation models (BDIM) \cite{koonin2002structure, wojtowicz2007evolution, reed2004model, csHuros2006probabilistic, hahn2005estimating, gabetta2010gene}, but so far no study has included catastrophes in a multivariable system. In single-variable models, such as those for ecosystem carbon content, catastrophes were introduced to simulate drastic changes in the environment, and studied using semi-stochastic models  \cite{leite2012stationary}. We show here that the main model results can be reproduced by a semi-stochastic model. The exponential model presented later, is very similar to the ecosystem carbon content model.

Formally, we study the effect of large-scale events (catastrophes) in a simple extension of the classical BDIM model, where the individuals belong to families defining different types, with the three  processes of the neutral BDIM:
\begin{enumerate}
  \item Birth: birth of an individual, the size of a certain family increases by 1.
  \item Death: death of an individual, the size of a certain family decreases by 1.
  \item Mutation: a constant fraction of all birth events leads to the emergence of new families. In effect, this is the creation of a new family of size $1$.
\end{enumerate}
To this, we add one more process.
\begin{enumerate}[resume]
  \item Catastrophe: a family is deleted and all individuals in this family are deleted from the system.
\end{enumerate}
In order to equilibrate the total population size, we assume that the death rate is proportional to the total population size, as in the standard logistic model.

The birth and death rates are equal among  families. The catastrophe rate is equal for all families (i.e.  the probability that a family would die in a catastrophe  is not affected by its size).  Since catastrophes affect large and small families at the same rate, one could expect  catastrophes to induce a more homogeneous family-size distribution. We here show that the model's results are the opposite. The presence of catastrophes cause, in effect, a  larger variance of family-sizes.

Formally, we denote the size of each family, $k$, as the number of individuals in this family. The zero moment (Eq. 2), $m_0$, is the total number of families, and the first moment, $m_1$ is the total number of individuals over all families. $m_0$ and $m_1$ are not constant.
The four processes above can be computed using the following reactions: 1) A birth of an individual occurs at rate $\alpha$. 2) A death of an individual occurs at rate $\delta=\frac{ m_1}{\NEO}$,  $\NEO$ being some arbitrary number that would be the population size in equilibrium in the absence of catastrophes. 3) The fraction of mutations out of all birth events is $\mu$.
4) A catastrophe occurs at rate $\gamma$. $\alpha$, $\NEO$, $\mu$, and $\gamma$ are free parameters.

Technically, at every time step, the total number of individuals, $m_1$, is calculated and the death rate is set to be $\delta=\frac{ m_1}{\NEO}$.  Once $\delta$ is determined, $\alpha$, $\delta$ and $\gamma$ are normalized by their sum, and a process (birth, death or catastrophe) is chosen randomly, according to these relative probabilities.

Denoting the number of families of size $k$ by $N_k$, the master equations resulting from the four processes above are as follows (up to a time scaling, see Supplemental Material \cite{supp1} for equations derivation):
\begin{equation}
\left\{
  \begin{aligned}
    \frac{dN_1}{dt} =& m_1 \bigg[ \mu \alpha - \alpha(1-\mu)\frac{N_1}{m_1} \\
                     &+ \frac{1}{\NEO} \Big(- N_{1} + 2N_{2}\Big)-\frac{\gamma N_1}{m_0} \bigg]
    \\
    \frac{dN_k}{dt} =& m_1 \bigg[ \frac{\alpha(1-\mu)}{m_1}\Big((k-1)N_{k-1}-kN_{k}\Big)  \qquad \text{for }k>1\\
                     &\frac{1}{\NEO}\Big(-kN_{k} +(k+1)N_{k+1}\Big) - \frac {\gamma N_{k}}{m_0} \bigg]
  \end{aligned}
  \right.
  \label{eq:dNkdtdN1dt}
\end{equation}
In order to estimate the macroscopic dynamics described in \eqns \ref{eq:dNkdtdN1dt}, it is instructive to compute the moments of the distribution
\begin{equation}
m_j=\sum_k k^jN_k,  \ j=0,1,2,
\label{eq:defmj}
\end{equation}
where $j$ is the moment order.
Substituting \eqns \ref{eq:dNkdtdN1dt} into \eqns \ref{eq:defmj} and summing over $k$,  leads to \eqns \ref{eq:dmjdt} for the time derivative of the moments (see Supplemental Material \cite{supp1} for equations derivation):

\begin{equation}
\left\{
  \begin{aligned}
    \frac{dm_0}{dt} &=  m_1 \Big[ \mu \alpha - \frac{ N_1}{\NEO}- \gamma \Big]\\
    \\
    \frac{dm_1}{dt} &= m_1 \Big[ \alpha  - \frac{ m_1}{\NEO} - \gamma\frac{m_1}{m_0} \Big]
    \\
    \frac{dm_2}{dt} &=  m_1 \bigg[ \alpha +\frac{m_1}{\NEO}\\
                    & +m_2\Big[\underbrace{ \frac{2\alpha(1-\mu)}{m_1}-\frac{2}{\NEO}- \frac{\gamma}{m_0} }_\text{B}\Big] \bigg]
  \end{aligned}
  \right.
  \label{eq:dmjdt}
\end{equation}

The $m_0$ and $m_1$ equations are closed and independent of $m_2$. They require, however, the value of $N_1$. The equation for $m_2$ is likewise exact and closed, requiring no higher order moments. This system of equations may be solved consistently if $N_1$  can be estimated.

In order to relate $N_1$ to the moments, one can  proceed with two additional assumptions: the continuous limit assumption and the scale free assumption.
The scale free assumption assumes that $N_k(k)$ has a scale free distribution,

\begin{equation}
N_k=N_1 k^{-\eta},
\label{eq:k_etta}
\end{equation}
where $\eta$ is the yet undetermined power. This is true for the model without catastrophes before the exponential cutoff, and is reasonable for low $k$ in the model with catastrophes. It is also  justified by a good fit between simulation and theory, as further discussed.

\eqn \ref{eq:k_etta} and the continuous limit assumption yield:
\begin{equation}
\left\{
  \begin{aligned}
    m_j &= \int_{1}^{\kmax}\!\!\!N_1 k^{j-\eta}dk = \frac{N_1(\kmax^{j+1-\eta}-1)}{j+1-\eta}, \qquad j=0,1,2
    \\
    N_2 &=N_1 2^{-\eta}.
  \end{aligned}
  \right.
  \label{eq:powerlawAssump}
\end{equation}

Substituting $N_2=N_12^{-\eta}$ into $\frac{dN_1}{dt}$ of \eqns \ref{eq:dNkdtdN1dt} leads to the steady-state equations:
\begin{equation}
\left\{
  \begin{aligned}
      N_1 &= \NEO\big( \mu \alpha-\gamma \big)
      \\
      m_0 &= \frac{\gamma N_1(2-\mu)}{\mu \alpha + \frac{N_1}{\NEO}(2^{1-\eta}-2+\mu)}
      \\
      m_1 &= \frac{ \alpha }{\frac{\gamma}{m_0} + \frac{1}{\NEO} }
      \\
      m_2 &= \frac{\alpha(\NEO\gamma +2m_0)}{(\NEO\gamma+m_0)\big(\frac{\gamma}{m_0}(2\mu-1)+\frac{2\mu}{\NEO}\big)}
 \end{aligned}
\right.
\label{eq:ss1}
\end{equation}

\begin{figure}
\includegraphics[width=0.40\textwidth]{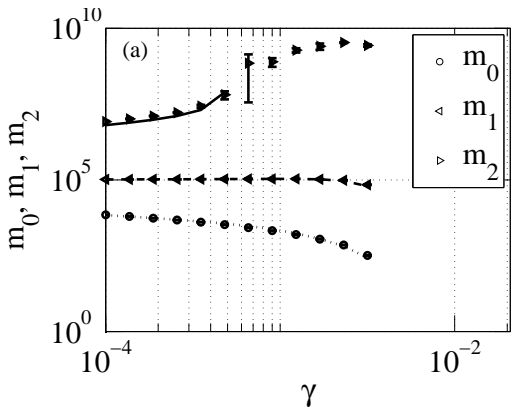}
\caption{Comparison of  simulation and theory. $m_0$ the number of families, $m_1$ the number of individuals, and $m_2$ the second moment.   Each point represents a dot on the black line of \fig \ref{fig:phaseDiagram}. See Supplemental Material \cite{supp4} for parameters. 
Each simulation result is the  average over $5$ realizations and $50$ time points separated by $10^6$ time steps after steady-state has been established. The lines are the theory and the dots are simulation. Theory and simulation are in perfect fit. $m_2$ diverges from point $7$ and on. The theoretic  $m_2$ in this regime is negative and is not shown here. Empty regions represent values of $0$. See \cite{supp4} for similar results for $N_1$. }
\label{fig:m0m1m2simTheory}
\end{figure}

An important result of the approximation of this relation when $\mu<<1$  is that $m_2$ is flat for values of $\gamma < \mu$, and increases with the value of $\gamma$ as $\gamma$ approaches $\mu$ (\fig \ref{fig:m0m1m2simTheory}). As  will be  further shown, when $\gamma > \mu$ the scale free assumption here fails, and a novel catastrophe induced  transition to extinction occurs (i.e. the total population collapses to 0).

The theoretical  solutions above are valid when:
\begin{equation}
\forall k,   N_k\geq 0 \Longrightarrow  \forall  j,  m_j\geq 0 .
\label{eq:positiveNk}
\end{equation}
 Imposing these conditions on the steady-state equations (\eqns \ref{eq:ss1}) leads to:
\begin{equation}
N_1  \geq 0 \Longrightarrow
\mu \alpha \geq \gamma
\label{eq:physicalPhase1}
\end{equation}
and
\begin{equation}
m_0  \geq 0 \Longrightarrow
D = \mu\alpha+\frac{N_1}{\NEO}(2^{1-\eta}-2+\mu) \geq 0.
\label{eq:physicalPhase2}
\end{equation}
The positivity of $m_1$ is trivial. The positivity of $m_2$ is  discussed  below.  The  phase transition  defined by both \eqn \ref{eq:physicalPhase1} and \eqn \ref{eq:physicalPhase2} represents the extinction-survival transition.

With no loss of generality, one can set $\alpha \approx 1$, leading to $m_1 \approx \NEO$, for low enough $\gamma$ values. In such a case, the first request resulting from the conditions above is  a mutation rate higher than the catastrophe rate, in order to balance the removed families by the creation of new families.
The second condition ensures a positive net entry of families to the $k=1$ family-size. If any of these two conditions is breached, the total population collapses.

When one considers the positivity of  $m_2$, a novel phase transition occurs. Both convergence and positivity of $m_2$ are determined by $B$, the pre-factor of $m_2$ in $\frac{dm_2}{dt}$ (\eqn \ref{eq:dmjdt}). $B<0$ is the condition for convergence and for positivity. When $B$ is positive the power-law assumption fails and the moment approximations  no longer hold. However, the power-law assumption is a good enough approximation to predict the location of the second phase transition, as the comparisons with simulation show. Substituting $m_1$ into $B$ we get
\begin{equation}
\gamma > \frac{2\mu}{1-2\mu}\frac{m_0}{\NEO}
\label{eq:gamma_c2}
\end{equation}
as the condition for both positivity and convergence.
In developing this equation we assumed that $\mu<\frac{1}{2}$, as extremely large mutation rates lead to a population collapse \cite{eigen2002error}.

Thus, a critical transition emerges, where the family-size variance diverges. This new second phase transition divides the survival phase into ``low variance" and ``high variance" phases.  The diversity of the family-size distribution is governed by $m_2$, and when $B$ becomes positive  one can expect high diversity in family-size over time and realizations. With the increasing diversity in family-size comes an increasing sensitivity to the collapse of very large families and the resulting fluctuating dynamics. See Supplemental Material \cite{supp4} for simulated $m_0$, $m_1$, and $m_2$ as a function of time. 
The differences between the dynamics on the two sides of the phase transition are very clear.

\begin{figure}
\includegraphics[width=0.40\textwidth]{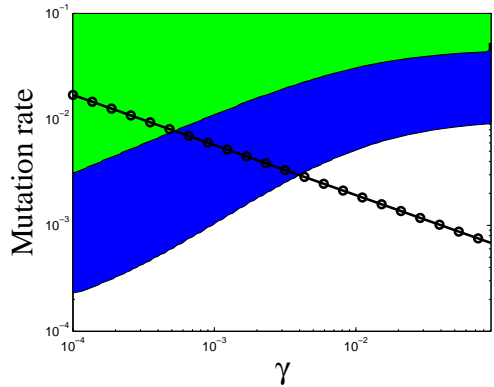}
\caption{Phase diagram. ``Low diversity" phase were $m_2$ convergence (green), ``high diversity" phase (blue) and extinction phase (white). Results of the theoretic model and solved numerically. Points on the black line were simulated and compared with the numeric solution (see \fig \ref{fig:m0m1m2simTheory}). }
\label{fig:phaseDiagram}
\end{figure}

The different phases of the model are summarized in the $2$-dimensional phase diagram, \fig \ref{fig:phaseDiagram}, which is a numeric solution of the first two equations of \eqns \ref{eq:powerlawAssump} and the first three equations of \eqns \ref{eq:ss1}. It denotes the extinction phase in white, and the line separating it from the survival phase is given by \eqn \ref{eq:physicalPhase1}. \eqn \ref{eq:physicalPhase2} is automatically satisfied when \eqn \ref{eq:physicalPhase1} is satisfied. The ``high variance" phase defined by \eqn \ref{eq:gamma_c2} is denoted in blue. The green area is the ``low variance" phase, where $B<0$.

One is thus led to the surprising result that as $\gamma$ increases beyond some value the diversity diverges, and the model above fails. Moreover, even in the domain where the model assumptions hold, $m_2$ grows with $\gamma$. Thus, in contradiction with current concepts, in the current model, increasing the catastrophe rate actually increases the diversity in time and in the family-size distribution. When $\gamma$ further grows to kill more families than are produced by mutations, the total population collapses, and extinction occurs through a novel phase transition.

\begin{figure}
\includegraphics[width=0.40\textwidth]{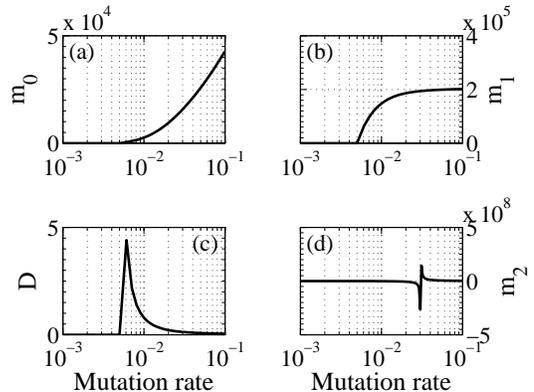}
\caption{ Numeric solution showing both phase transitions. The first phase transition from extinction to survival phase is evident in $D$, the denominator of $m_0$. The second phase transition, from ``high diversity" to ``low diversity", is evident in $m_2$. Here $\gamma=0.01$, $\alpha=1.99$, and $\NEO=103890$. }
\label{fig:m0Theory}
\end{figure}

Both phase transitions are further illustrated in \fig \ref{fig:m0Theory}, where the theoretic solution for the moments and the denominator of $m_0$ is given, with fixed $\alpha$, $\gamma$, and $\NEO$, but changing $\mu$. For small $\mu$ values the system is extinct.  A first phase transition occurs  at $\mu\simeq0.006$, where the system enters  the ``high variance" phase, and although $m_0$ and $m_1$ are positive, $m_2$  would be negative in the scale free approximation. After $\mu\simeq0.03$,  $m_2$, becomes  positive too.

\begin{figure}
\includegraphics[width=0.40\textwidth]{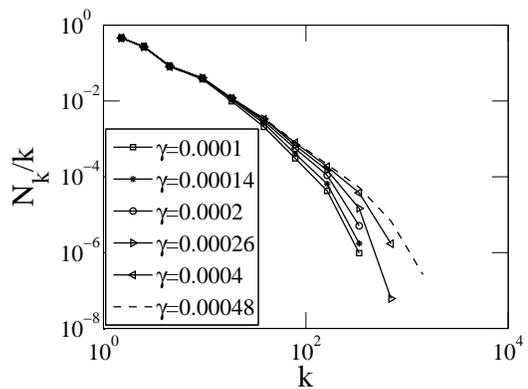}
\caption{Simulated steady-state family-size distribution, $N_k/k$. A line represents a system with parameters defined by dots on the black line of \fig \ref{fig:phaseDiagram} (see Supplemental Material \cite{supp4} for parameters). 
As $\gamma$ increases the number of catastrophes increases, the relative number of larger families increases. The maximal family size increases as well. }
\label{fig:NkSim}
\end{figure}

Simulation results confirming the theoretic solution are given in \fig  \ref{fig:m0m1m2simTheory}. See Supplemental Material \cite{supp4} for the set of parameters corresponding to each dot in \fig \ref{fig:m0m1m2simTheory}, \fig \ref{fig:phaseDiagram}, and \fig \ref{fig:NkSim}. We chose a set of systems sampling all phases of the model. The first six dots are systems in the ``low variance" phase, the next six dots are systems in the ``high variance" phase and the remaining ten dots are systems in the extinction  phase. The  theoretic solution is valid   in the ``low variance" phase, where a very good agreement with  simulations can be observed. In the ``high variance" phase, a good agreement with simulations is observed  for all moments except $m_2$ . The ten extinction phase simulations have very quickly converged with all moments reaching zero, and are not plotted in the graphs.

The increase in $m_2$ is also clear in the  simulated steady-state family-size distribution. The steady-state size distributions of systems corresponding to the dots on the black line are given in  \fig \ref{fig:NkSim}. As $\gamma$ increases, the slope of the power-law decreases, the fraction of large-size families increases, and  the maximal family-size increases.

The intuition behind this mechanism is simple. In the absence of catastrophes, births and deaths must be balanced leading to a Ewens-like family-size distribution \cite{ewens1972sampling}.  However, in the presence of catastrophes, the average birth rate of families can be higher than the death rate, with the total population balanced by catastrophes. In such a regime, with an average net growth rate of $\theta$, the family size of a family $k$ is on average as $e^{\theta \tau_k}$, where $\tau_k$ is the time from the emergence of this family through a mutation to the current time, unless it was destroyed by a catastrophe. For large enough values of $\mu$ the total number of families is arbitrarily large, and the time between catastrophes for a given family increases linearly with the number of families.

In order to validate that such a model does produce the increase in diversity following the increase in catastrophes rate, we simulated a model where in each time step $\Delta t$, the following reactions occur:
\begin{itemize}
\item $\gamma m_0$ out of $m_0$ families are destroyed using a random choice with a Binomial distribution.
\item $\mu m_1$ families are produced randomly with a Poisson distribution.
\item Each family grows by a factor of $exp\big[(\alpha-\mu-m_1/N)\Delta t\big]$.
 \end{itemize}
 Here, as in the original model, $\alpha$ is the birth rate, $N$ is some arbitrary large number, $\mu$ is the mutation rate, and $\gamma$ is the catastrophes rate. Indeed, this simulation reproduces the relation above (see Supplemental Material \cite{supp5} for family-size distribution of this model), 
 with the clear phase transition to very high $m_2$ values. Moreover, the catastrophe size increases with $\gamma$. When $\gamma$ becomes larger than $\mu$, the total population collapses as in the original model.

While we have here studied a single specific model, the same results hold in different domains, with different dynamics (see Supplemental Material \cite{supp2, supp3} for details of other models where the same results hold). The main element driving these results is that, in equilibrium, the total population is governed by a balance between family growth and decay. However, while the death term affects all families equally, the catastrophe term affects a random subset of families.
Since a part of the death term is balanced by the catastrophes, the remaining total death term is lower than the total birth rate. Thus, for families not dying by catastrophes, the net difference between growth and death is positive, leading to an exponential growth, until a catastrophe occurs. Therefore, increasing the catastrophe rate increases the net growth rate and decreases the time between catastrophe events.
Such a balance can be observed in many scenarios, such as the growth of stock values in stock markets, family growth in highly fluctuating environments, or the dynamics of network, where vertices can accumulate edges, until a vertex deletion event happens.

While large jumps in single families (e.g. L\'evy flights) have been extensively studied \cite{Levyflights}, these jumps are assumed to be averaged and not to induce very large changes in  large ensembles. This is in clear contrast with large-scale fluctuations observed, among others, in the total market value of stock markets. The influence of frequent catastrophes may be one element explaining these fluctuations. An important element obviously missing from the current description is the interaction between families and the possibility of cascades from one collapsing population to the other. We now plan to study whether the mechanisms described here apply when interactions are taken into account.

\bibliographystyle{apsrev4-1}

\begin{thebibliography}{44}%
\makeatletter
\providecommand \@ifxundefined [1]{%
 \@ifx{#1\undefined}
}%
\providecommand \@ifnum [1]{%
 \ifnum #1\expandafter \@firstoftwo
 \else \expandafter \@secondoftwo
 \fi
}%
\providecommand \@ifx [1]{%
 \ifx #1\expandafter \@firstoftwo
 \else \expandafter \@secondoftwo
 \fi
}%
\providecommand \natexlab [1]{#1}%
\providecommand \enquote  [1]{``#1''}%
\providecommand \bibnamefont  [1]{#1}%
\providecommand \bibfnamefont [1]{#1}%
\providecommand \citenamefont [1]{#1}%
\providecommand \href@noop [0]{\@secondoftwo}%
\providecommand \href [0]{\begingroup \@sanitize@url \@href}%
\providecommand \@href[1]{\@@startlink{#1}\@@href}%
\providecommand \@@href[1]{\endgroup#1\@@endlink}%
\providecommand \@sanitize@url [0]{\catcode `\\12\catcode `\$12\catcode
  `\&12\catcode `\#12\catcode `\^12\catcode `\_12\catcode `\%12\relax}%
\providecommand \@@startlink[1]{}%
\providecommand \@@endlink[0]{}%
\providecommand \url  [0]{\begingroup\@sanitize@url \@url }%
\providecommand \@url [1]{\endgroup\@href {#1}{\urlprefix }}%
\providecommand \urlprefix  [0]{URL }%
\providecommand \Eprint [0]{\href }%
\providecommand \doibase [0]{http://dx.doi.org/}%
\providecommand \selectlanguage [0]{\@gobble}%
\providecommand \bibinfo  [0]{\@secondoftwo}%
\providecommand \bibfield  [0]{\@secondoftwo}%
\providecommand \translation [1]{[#1]}%
\providecommand \BibitemOpen [0]{}%
\providecommand \bibitemStop [0]{}%
\providecommand \bibitemNoStop [0]{.\EOS\space}%
\providecommand \EOS [0]{\spacefactor3000\relax}%
\providecommand \BibitemShut  [1]{\csname bibitem#1\endcsname}%
\let\auto@bib@innerbib\@empty
\bibitem [{\citenamefont {Weng}\ \emph {et~al.}(2012)\citenamefont {Weng},
  \citenamefont {Luo}, \citenamefont {Wang}, \citenamefont {Wang},
  \citenamefont {Hayes}, \citenamefont {McGuire}, \citenamefont {Hastings},\
  and\ \citenamefont {Schimel}}]{weng2012ecosystem}%
  \BibitemOpen
  \bibfield  {author} {\bibinfo {author} {\bibfnamefont {E.}~\bibnamefont
  {Weng}}, \bibinfo {author} {\bibfnamefont {Y.}~\bibnamefont {Luo}}, \bibinfo
  {author} {\bibfnamefont {W.}~\bibnamefont {Wang}}, \bibinfo {author}
  {\bibfnamefont {H.}~\bibnamefont {Wang}}, \bibinfo {author} {\bibfnamefont
  {D.~J.}\ \bibnamefont {Hayes}}, \bibinfo {author} {\bibfnamefont {A.~D.}\
  \bibnamefont {McGuire}}, \bibinfo {author} {\bibfnamefont {A.}~\bibnamefont
  {Hastings}}, \ and\ \bibinfo {author} {\bibfnamefont {D.~S.}\ \bibnamefont
  {Schimel}},\ }\href@noop {} {\bibfield  {journal} {\bibinfo  {journal}
  {Journal of Geophysical Research: Biogeosciences}\ }\textbf {\bibinfo
  {volume} {117}} (\bibinfo {year} {2012})}\BibitemShut {NoStop}%
\bibitem [{\citenamefont {Leite}\ \emph {et~al.}(2012)\citenamefont {Leite},
  \citenamefont {Petrov},\ and\ \citenamefont {Weng}}]{leite2012stationary}%
  \BibitemOpen
  \bibfield  {author} {\bibinfo {author} {\bibfnamefont {M.~C.~A.}\
  \bibnamefont {Leite}}, \bibinfo {author} {\bibfnamefont {N.~P.}\ \bibnamefont
  {Petrov}}, \ and\ \bibinfo {author} {\bibfnamefont {E.}~\bibnamefont
  {Weng}},\ }\href@noop {} {\bibfield  {journal} {\bibinfo  {journal}
  {Nonlinear Analysis: Real World Applications}\ }\textbf {\bibinfo {volume}
  {13}},\ \bibinfo {pages} {497} (\bibinfo {year} {2012})}\BibitemShut
  {NoStop}%
\bibitem [{\citenamefont {Mawby}\ \emph {et~al.}(1989)\citenamefont {Mawby},
  \citenamefont {Hain},\ and\ \citenamefont {Doggett}}]{mawby1989endemic}%
  \BibitemOpen
  \bibfield  {author} {\bibinfo {author} {\bibfnamefont {W.}~\bibnamefont
  {Mawby}}, \bibinfo {author} {\bibfnamefont {F.}~\bibnamefont {Hain}}, \ and\
  \bibinfo {author} {\bibfnamefont {C.}~\bibnamefont {Doggett}},\ }\href@noop
  {} {\bibfield  {journal} {\bibinfo  {journal} {Forest Science}\ }\textbf
  {\bibinfo {volume} {35}},\ \bibinfo {pages} {1075} (\bibinfo {year}
  {1989})}\BibitemShut {NoStop}%
\bibitem [{\citenamefont {Turcotte}(1999)}]{turcotte1999self}%
  \BibitemOpen
  \bibfield  {author} {\bibinfo {author} {\bibfnamefont {D.~L.}\ \bibnamefont
  {Turcotte}},\ }\href@noop {} {\bibfield  {journal} {\bibinfo  {journal}
  {Reports on Progress in Physics}\ }\textbf {\bibinfo {volume} {62}},\
  \bibinfo {pages} {1377} (\bibinfo {year} {1999})}\BibitemShut {NoStop}%
\bibitem [{\citenamefont {Chen-Charpentier}\ and\ \citenamefont
  {Leite}(2014)}]{chen2014model}%
  \BibitemOpen
  \bibfield  {author} {\bibinfo {author} {\bibfnamefont {B.}~\bibnamefont
  {Chen-Charpentier}}\ and\ \bibinfo {author} {\bibfnamefont {M.}~\bibnamefont
  {Leite}},\ }\href@noop {} {\bibfield  {journal} {\bibinfo  {journal}
  {Ecological Modelling}\ }\textbf {\bibinfo {volume} {286}},\ \bibinfo {pages}
  {26} (\bibinfo {year} {2014})}\BibitemShut {NoStop}%
\bibitem [{\citenamefont {Wilcox}\ and\ \citenamefont
  {Elderd}(2003)}]{wilcox2003effect}%
  \BibitemOpen
  \bibfield  {author} {\bibinfo {author} {\bibfnamefont {C.}~\bibnamefont
  {Wilcox}}\ and\ \bibinfo {author} {\bibfnamefont {B.}~\bibnamefont
  {Elderd}},\ }\href@noop {} {\bibfield  {journal} {\bibinfo  {journal}
  {Journal of Applied Ecology}\ }\textbf {\bibinfo {volume} {40}},\ \bibinfo
  {pages} {859} (\bibinfo {year} {2003})}\BibitemShut {NoStop}%
\bibitem [{\citenamefont {Petraitis}(2013)}]{petraitis2013multiple}%
  \BibitemOpen
  \bibfield  {author} {\bibinfo {author} {\bibfnamefont {P.}~\bibnamefont
  {Petraitis}},\ }\href@noop {} {\emph {\bibinfo {title} {Multiple stable
  states in natural ecosystems}}}\ (\bibinfo  {publisher} {OUP Oxford},\
  \bibinfo {year} {2013})\BibitemShut {NoStop}%
\bibitem [{\citenamefont {Scheffer}\ \emph {et~al.}(2001)\citenamefont
  {Scheffer}, \citenamefont {Carpenter}, \citenamefont {Foley}, \citenamefont
  {Folke},\ and\ \citenamefont {Walker}}]{scheffer2001catastrophic}%
  \BibitemOpen
  \bibfield  {author} {\bibinfo {author} {\bibfnamefont {M.}~\bibnamefont
  {Scheffer}}, \bibinfo {author} {\bibfnamefont {S.}~\bibnamefont {Carpenter}},
  \bibinfo {author} {\bibfnamefont {J.~A.}\ \bibnamefont {Foley}}, \bibinfo
  {author} {\bibfnamefont {C.}~\bibnamefont {Folke}}, \ and\ \bibinfo {author}
  {\bibfnamefont {B.}~\bibnamefont {Walker}},\ }\href@noop {} {\bibfield
  {journal} {\bibinfo  {journal} {Nature}\ }\textbf {\bibinfo {volume} {413}},\
  \bibinfo {pages} {591} (\bibinfo {year} {2001})}\BibitemShut {NoStop}%
\bibitem [{\citenamefont {Ives}\ and\ \citenamefont
  {Cardinale}(2004)}]{ives2004food}%
  \BibitemOpen
  \bibfield  {author} {\bibinfo {author} {\bibfnamefont {A.~R.}\ \bibnamefont
  {Ives}}\ and\ \bibinfo {author} {\bibfnamefont {B.~J.}\ \bibnamefont
  {Cardinale}},\ }\href@noop {} {\bibfield  {journal} {\bibinfo  {journal}
  {Nature}\ }\textbf {\bibinfo {volume} {429}},\ \bibinfo {pages} {174}
  (\bibinfo {year} {2004})}\BibitemShut {NoStop}%
\bibitem [{\citenamefont {Hart}\ and\ \citenamefont
  {Avil{\'e}s}(2014)}]{hart2014reconstructing}%
  \BibitemOpen
  \bibfield  {author} {\bibinfo {author} {\bibfnamefont {E.~M.}\ \bibnamefont
  {Hart}}\ and\ \bibinfo {author} {\bibfnamefont {L.}~\bibnamefont
  {Avil{\'e}s}},\ }\href@noop {} {\bibfield  {journal} {\bibinfo  {journal}
  {PloS one}\ }\textbf {\bibinfo {volume} {9}},\ \bibinfo {pages} {e110049}
  (\bibinfo {year} {2014})}\BibitemShut {NoStop}%
\bibitem [{\citenamefont {Tuljapurkar}(2013)}]{tuljapurkar2013population}%
  \BibitemOpen
  \bibfield  {author} {\bibinfo {author} {\bibfnamefont {S.}~\bibnamefont
  {Tuljapurkar}},\ }\href@noop {} {\emph {\bibinfo {title} {Population dynamics
  in variable environments}}},\ Vol.~\bibinfo {volume} {85}\ (\bibinfo
  {publisher} {Springer Science \& Business Media},\ \bibinfo {year}
  {2013})\BibitemShut {NoStop}%
\bibitem [{\citenamefont {Froot}\ and\ \citenamefont
  {O’Connell}(2008)}]{froot2008pricing}%
  \BibitemOpen
  \bibfield  {author} {\bibinfo {author} {\bibfnamefont {K.~A.}\ \bibnamefont
  {Froot}}\ and\ \bibinfo {author} {\bibfnamefont {P.~G.}\ \bibnamefont
  {O’Connell}},\ }\href@noop {} {\bibfield  {journal} {\bibinfo  {journal}
  {Journal of Banking \& Finance}\ }\textbf {\bibinfo {volume} {32}},\ \bibinfo
  {pages} {69} (\bibinfo {year} {2008})}\BibitemShut {NoStop}%
\bibitem [{\citenamefont {He}\ and\ \citenamefont
  {Krishnamurthy}(2012)}]{he2012model}%
  \BibitemOpen
  \bibfield  {author} {\bibinfo {author} {\bibfnamefont {Z.}~\bibnamefont
  {He}}\ and\ \bibinfo {author} {\bibfnamefont {A.}~\bibnamefont
  {Krishnamurthy}},\ }\href@noop {} {\bibfield  {journal} {\bibinfo  {journal}
  {The Review of Economic Studies}\ }\textbf {\bibinfo {volume} {79}},\
  \bibinfo {pages} {735} (\bibinfo {year} {2012})}\BibitemShut {NoStop}%
\bibitem [{\citenamefont {Enjolras}\ and\ \citenamefont
  {Kast}(2012)}]{enjolras2012combining}%
  \BibitemOpen
  \bibfield  {author} {\bibinfo {author} {\bibfnamefont {G.}~\bibnamefont
  {Enjolras}}\ and\ \bibinfo {author} {\bibfnamefont {R.}~\bibnamefont
  {Kast}},\ }\href@noop {} {\bibfield  {journal} {\bibinfo  {journal}
  {Agricultural Finance Review}\ }\textbf {\bibinfo {volume} {72}},\ \bibinfo
  {pages} {156} (\bibinfo {year} {2012})}\BibitemShut {NoStop}%
\bibitem [{\citenamefont {Nowak}\ and\ \citenamefont
  {Romaniuk}(2013)}]{nowak2013pricing}%
  \BibitemOpen
  \bibfield  {author} {\bibinfo {author} {\bibfnamefont {P.}~\bibnamefont
  {Nowak}}\ and\ \bibinfo {author} {\bibfnamefont {M.}~\bibnamefont
  {Romaniuk}},\ }\href@noop {} {\bibfield  {journal} {\bibinfo  {journal}
  {Insurance: Mathematics and Economics}\ }\textbf {\bibinfo {volume} {52}},\
  \bibinfo {pages} {18} (\bibinfo {year} {2013})}\BibitemShut {NoStop}%
\bibitem [{\citenamefont {Falco}\ \emph {et~al.}(2014)\citenamefont {Falco},
  \citenamefont {Adinolfi}, \citenamefont {Bozzola},\ and\ \citenamefont
  {Capitanio}}]{falco2014crop}%
  \BibitemOpen
  \bibfield  {author} {\bibinfo {author} {\bibfnamefont {S.~D.}\ \bibnamefont
  {Falco}}, \bibinfo {author} {\bibfnamefont {F.}~\bibnamefont {Adinolfi}},
  \bibinfo {author} {\bibfnamefont {M.}~\bibnamefont {Bozzola}}, \ and\
  \bibinfo {author} {\bibfnamefont {F.}~\bibnamefont {Capitanio}},\ }\href@noop
  {} {\bibfield  {journal} {\bibinfo  {journal} {Journal of Agricultural
  Economics}\ }\textbf {\bibinfo {volume} {65}},\ \bibinfo {pages} {485}
  (\bibinfo {year} {2014})}\BibitemShut {NoStop}%
\bibitem [{\citenamefont {Pruessner}(2012)}]{pruessner2012self}%
  \BibitemOpen
  \bibfield  {author} {\bibinfo {author} {\bibfnamefont {G.}~\bibnamefont
  {Pruessner}},\ }\href@noop {} {\emph {\bibinfo {title} {Self-organised
  criticality: theory, models and characterisation}}}\ (\bibinfo  {publisher}
  {Cambridge University Press},\ \bibinfo {year} {2012})\BibitemShut {NoStop}%
\bibitem [{\citenamefont {Malamud}\ \emph {et~al.}(1998)\citenamefont
  {Malamud}, \citenamefont {Morein},\ and\ \citenamefont
  {Turcotte}}]{malamud1998forest}%
  \BibitemOpen
  \bibfield  {author} {\bibinfo {author} {\bibfnamefont {B.~D.}\ \bibnamefont
  {Malamud}}, \bibinfo {author} {\bibfnamefont {G.}~\bibnamefont {Morein}}, \
  and\ \bibinfo {author} {\bibfnamefont {D.~L.}\ \bibnamefont {Turcotte}},\
  }\href@noop {} {\bibfield  {journal} {\bibinfo  {journal} {Science}\ }\textbf
  {\bibinfo {volume} {281}},\ \bibinfo {pages} {1840} (\bibinfo {year}
  {1998})}\BibitemShut {NoStop}%
\bibitem [{\citenamefont {Drossel}\ and\ \citenamefont
  {Schwabl}(1992)}]{drossel1992self}%
  \BibitemOpen
  \bibfield  {author} {\bibinfo {author} {\bibfnamefont {B.}~\bibnamefont
  {Drossel}}\ and\ \bibinfo {author} {\bibfnamefont {F.}~\bibnamefont
  {Schwabl}},\ }\href@noop {} {\bibfield  {journal} {\bibinfo  {journal}
  {Physical Review Letters}\ }\textbf {\bibinfo {volume} {69}},\ \bibinfo
  {pages} {1629} (\bibinfo {year} {1992})}\BibitemShut {NoStop}%
\bibitem [{\citenamefont {Bak}\ \emph {et~al.}(1988)\citenamefont {Bak},
  \citenamefont {Tang},\ and\ \citenamefont {Wiesenfeld}}]{bak1988self}%
  \BibitemOpen
  \bibfield  {author} {\bibinfo {author} {\bibfnamefont {P.}~\bibnamefont
  {Bak}}, \bibinfo {author} {\bibfnamefont {C.}~\bibnamefont {Tang}}, \ and\
  \bibinfo {author} {\bibfnamefont {K.}~\bibnamefont {Wiesenfeld}},\
  }\href@noop {} {\bibfield  {journal} {\bibinfo  {journal} {Physical Review
  A}\ }\textbf {\bibinfo {volume} {38}},\ \bibinfo {pages} {364} (\bibinfo
  {year} {1988})}\BibitemShut {NoStop}%
\bibitem [{\citenamefont {Kadanoff}\ \emph {et~al.}(1989)\citenamefont
  {Kadanoff}, \citenamefont {Nagel}, \citenamefont {Wu},\ and\ \citenamefont
  {Zhou}}]{kadanoff1989scaling}%
  \BibitemOpen
  \bibfield  {author} {\bibinfo {author} {\bibfnamefont {L.~P.}\ \bibnamefont
  {Kadanoff}}, \bibinfo {author} {\bibfnamefont {S.~R.}\ \bibnamefont {Nagel}},
  \bibinfo {author} {\bibfnamefont {L.}~\bibnamefont {Wu}}, \ and\ \bibinfo
  {author} {\bibfnamefont {S.-m.}\ \bibnamefont {Zhou}},\ }\href@noop {}
  {\bibfield  {journal} {\bibinfo  {journal} {Physical Review A}\ }\textbf
  {\bibinfo {volume} {39}},\ \bibinfo {pages} {6524} (\bibinfo {year}
  {1989})}\BibitemShut {NoStop}%
\bibitem [{\citenamefont {Yoder}\ \emph {et~al.}(2011)\citenamefont {Yoder},
  \citenamefont {Turcotte},\ and\ \citenamefont {Rundle}}]{yoder2011forest}%
  \BibitemOpen
  \bibfield  {author} {\bibinfo {author} {\bibfnamefont {M.~R.}\ \bibnamefont
  {Yoder}}, \bibinfo {author} {\bibfnamefont {D.~L.}\ \bibnamefont {Turcotte}},
  \ and\ \bibinfo {author} {\bibfnamefont {J.~B.}\ \bibnamefont {Rundle}},\
  }\href@noop {} {\bibfield  {journal} {\bibinfo  {journal} {Physical Review
  E}\ }\textbf {\bibinfo {volume} {83}},\ \bibinfo {pages} {046118} (\bibinfo
  {year} {2011})}\BibitemShut {NoStop}%
\bibitem [{\citenamefont {Clar}\ \emph {et~al.}(1994)\citenamefont {Clar},
  \citenamefont {Drossel},\ and\ \citenamefont {Schwabl}}]{clar1994scaling}%
  \BibitemOpen
  \bibfield  {author} {\bibinfo {author} {\bibfnamefont {S.}~\bibnamefont
  {Clar}}, \bibinfo {author} {\bibfnamefont {B.}~\bibnamefont {Drossel}}, \
  and\ \bibinfo {author} {\bibfnamefont {F.}~\bibnamefont {Schwabl}},\
  }\href@noop {} {\bibfield  {journal} {\bibinfo  {journal} {Physical Review
  E}\ }\textbf {\bibinfo {volume} {50}},\ \bibinfo {pages} {1009} (\bibinfo
  {year} {1994})}\BibitemShut {NoStop}%
\bibitem [{\citenamefont {Clar}\ \emph {et~al.}(1996)\citenamefont {Clar},
  \citenamefont {Drossel},\ and\ \citenamefont {Schwabl}}]{clar1996forest}%
  \BibitemOpen
  \bibfield  {author} {\bibinfo {author} {\bibfnamefont {S.}~\bibnamefont
  {Clar}}, \bibinfo {author} {\bibfnamefont {B.}~\bibnamefont {Drossel}}, \
  and\ \bibinfo {author} {\bibfnamefont {F.}~\bibnamefont {Schwabl}},\
  }\href@noop {} {\bibfield  {journal} {\bibinfo  {journal} {Journal of
  Physics: Condensed Matter}\ }\textbf {\bibinfo {volume} {8}},\ \bibinfo
  {pages} {6803} (\bibinfo {year} {1996})}\BibitemShut {NoStop}%
\bibitem [{\citenamefont {No{\"e}l}\ \emph {et~al.}(2013)\citenamefont
  {No{\"e}l}, \citenamefont {Brummitt},\ and\ \citenamefont
  {D’Souza}}]{noel2013controlling}%
  \BibitemOpen
  \bibfield  {author} {\bibinfo {author} {\bibfnamefont {P.-A.}\ \bibnamefont
  {No{\"e}l}}, \bibinfo {author} {\bibfnamefont {C.~D.}\ \bibnamefont
  {Brummitt}}, \ and\ \bibinfo {author} {\bibfnamefont {R.~M.}\ \bibnamefont
  {D’Souza}},\ }\href@noop {} {\bibfield  {journal} {\bibinfo  {journal}
  {Physical Review Letters}\ }\textbf {\bibinfo {volume} {111}},\ \bibinfo
  {pages} {078701} (\bibinfo {year} {2013})}\BibitemShut {NoStop}%
\bibitem [{\citenamefont {Stanley}\ \emph {et~al.}(2002)\citenamefont
  {Stanley}, \citenamefont {Amaral}, \citenamefont {Gopikrishnan},
  \citenamefont {Plerou},\ and\ \citenamefont
  {Rosenow}}]{stanley2002quantifying}%
  \BibitemOpen
  \bibfield  {author} {\bibinfo {author} {\bibfnamefont {H.~E.}\ \bibnamefont
  {Stanley}}, \bibinfo {author} {\bibfnamefont {L.~A.~N.}\ \bibnamefont
  {Amaral}}, \bibinfo {author} {\bibfnamefont {P.}~\bibnamefont
  {Gopikrishnan}}, \bibinfo {author} {\bibfnamefont {V.}~\bibnamefont
  {Plerou}}, \ and\ \bibinfo {author} {\bibfnamefont {B.}~\bibnamefont
  {Rosenow}},\ }in\ \href@noop {} {\emph {\bibinfo {booktitle} {Empirical
  Science of Financial Fluctuations}}}\ (\bibinfo  {publisher} {Springer},\
  \bibinfo {year} {2002})\ pp.\ \bibinfo {pages} {3--11}\BibitemShut {NoStop}%
\bibitem [{\citenamefont {Fu}\ \emph {et~al.}(2005)\citenamefont {Fu},
  \citenamefont {Pammolli}, \citenamefont {Buldyrev}, \citenamefont
  {Riccaboni}, \citenamefont {Matia}, \citenamefont {Yamasaki},\ and\
  \citenamefont {Stanley}}]{fu2005growth}%
  \BibitemOpen
  \bibfield  {author} {\bibinfo {author} {\bibfnamefont {D.}~\bibnamefont
  {Fu}}, \bibinfo {author} {\bibfnamefont {F.}~\bibnamefont {Pammolli}},
  \bibinfo {author} {\bibfnamefont {S.~V.}\ \bibnamefont {Buldyrev}}, \bibinfo
  {author} {\bibfnamefont {M.}~\bibnamefont {Riccaboni}}, \bibinfo {author}
  {\bibfnamefont {K.}~\bibnamefont {Matia}}, \bibinfo {author} {\bibfnamefont
  {K.}~\bibnamefont {Yamasaki}}, \ and\ \bibinfo {author} {\bibfnamefont
  {H.~E.}\ \bibnamefont {Stanley}},\ }\href@noop {} {\bibfield  {journal}
  {\bibinfo  {journal} {Proceedings of the National Academy of Sciences of the
  United States of America}\ }\textbf {\bibinfo {volume} {102}},\ \bibinfo
  {pages} {18801} (\bibinfo {year} {2005})}\BibitemShut {NoStop}%
\bibitem [{\citenamefont {Huynen}\ and\ \citenamefont
  {Van~Nimwegen}(1998)}]{huynen1998frequency}%
  \BibitemOpen
  \bibfield  {author} {\bibinfo {author} {\bibfnamefont {M.~A.}\ \bibnamefont
  {Huynen}}\ and\ \bibinfo {author} {\bibfnamefont {E.}~\bibnamefont
  {Van~Nimwegen}},\ }\href@noop {} {\bibfield  {journal} {\bibinfo  {journal}
  {Molecular Biology and Evolution}\ }\textbf {\bibinfo {volume} {15}},\
  \bibinfo {pages} {583} (\bibinfo {year} {1998})}\BibitemShut {NoStop}%
\bibitem [{\citenamefont {Koonin}\ \emph {et~al.}(2002)\citenamefont {Koonin},
  \citenamefont {Wolf},\ and\ \citenamefont {Karev}}]{koonin2002structure}%
  \BibitemOpen
  \bibfield  {author} {\bibinfo {author} {\bibfnamefont {E.~V.}\ \bibnamefont
  {Koonin}}, \bibinfo {author} {\bibfnamefont {Y.~I.}\ \bibnamefont {Wolf}}, \
  and\ \bibinfo {author} {\bibfnamefont {G.~P.}\ \bibnamefont {Karev}},\
  }\href@noop {} {\bibfield  {journal} {\bibinfo  {journal} {Nature}\ }\textbf
  {\bibinfo {volume} {420}},\ \bibinfo {pages} {218} (\bibinfo {year}
  {2002})}\BibitemShut {NoStop}%
\bibitem [{\citenamefont {Qian}\ \emph {et~al.}(2001)\citenamefont {Qian},
  \citenamefont {Luscombe},\ and\ \citenamefont {Gerstein}}]{qian2001protein}%
  \BibitemOpen
  \bibfield  {author} {\bibinfo {author} {\bibfnamefont {J.}~\bibnamefont
  {Qian}}, \bibinfo {author} {\bibfnamefont {N.~M.}\ \bibnamefont {Luscombe}},
  \ and\ \bibinfo {author} {\bibfnamefont {M.}~\bibnamefont {Gerstein}},\
  }\href@noop {} {\bibfield  {journal} {\bibinfo  {journal} {Journal of
  Molecular Biology}\ }\textbf {\bibinfo {volume} {313}},\ \bibinfo {pages}
  {673} (\bibinfo {year} {2001})}\BibitemShut {NoStop}%
\bibitem [{\citenamefont {Luscombe}\ \emph {et~al.}(2002)\citenamefont
  {Luscombe}, \citenamefont {Qian}, \citenamefont {Zhang}, \citenamefont
  {Johnson},\ and\ \citenamefont {Gerstein}}]{luscombe2002dominance}%
  \BibitemOpen
  \bibfield  {author} {\bibinfo {author} {\bibfnamefont {N.~M.}\ \bibnamefont
  {Luscombe}}, \bibinfo {author} {\bibfnamefont {J.}~\bibnamefont {Qian}},
  \bibinfo {author} {\bibfnamefont {Z.}~\bibnamefont {Zhang}}, \bibinfo
  {author} {\bibfnamefont {T.}~\bibnamefont {Johnson}}, \ and\ \bibinfo
  {author} {\bibfnamefont {M.}~\bibnamefont {Gerstein}},\ }\href@noop {}
  {\bibfield  {journal} {\bibinfo  {journal} {Genome Biology}\ }\textbf
  {\bibinfo {volume} {3}},\ \bibinfo {pages} {Research0040} (\bibinfo {year}
  {2002})}\BibitemShut {NoStop}%
\bibitem [{\citenamefont {W{\'o}jtowicz}\ and\ \citenamefont
  {Tiuryn}(2007)}]{wojtowicz2007evolution}%
  \BibitemOpen
  \bibfield  {author} {\bibinfo {author} {\bibfnamefont {D.}~\bibnamefont
  {W{\'o}jtowicz}}\ and\ \bibinfo {author} {\bibfnamefont {J.}~\bibnamefont
  {Tiuryn}},\ }\href@noop {} {\bibfield  {journal} {\bibinfo  {journal}
  {Journal of Computational Biology}\ }\textbf {\bibinfo {volume} {14}},\
  \bibinfo {pages} {479} (\bibinfo {year} {2007})}\BibitemShut {NoStop}%
\bibitem [{\citenamefont {Reed}\ and\ \citenamefont
  {Hughes}(2004)}]{reed2004model}%
  \BibitemOpen
  \bibfield  {author} {\bibinfo {author} {\bibfnamefont {W.~J.}\ \bibnamefont
  {Reed}}\ and\ \bibinfo {author} {\bibfnamefont {B.~D.}\ \bibnamefont
  {Hughes}},\ }\href@noop {} {\bibfield  {journal} {\bibinfo  {journal}
  {Mathematical Biosciences}\ }\textbf {\bibinfo {volume} {189}},\ \bibinfo
  {pages} {97} (\bibinfo {year} {2004})}\BibitemShut {NoStop}%
\bibitem [{\citenamefont {Cs{\H{u}}r{\"o}s}\ and\ \citenamefont
  {Mikl{\'o}s}(2006)}]{csHuros2006probabilistic}%
  \BibitemOpen
  \bibfield  {author} {\bibinfo {author} {\bibfnamefont {M.}~\bibnamefont
  {Cs{\H{u}}r{\"o}s}}\ and\ \bibinfo {author} {\bibfnamefont {I.}~\bibnamefont
  {Mikl{\'o}s}},\ }in\ \href@noop {} {\emph {\bibinfo {booktitle} {Annual
  International Conference on Research in Computational Molecular Biology}}}\
  (\bibinfo {organization} {Springer},\ \bibinfo {year} {2006})\ pp.\ \bibinfo
  {pages} {206--220}\BibitemShut {NoStop}%
\bibitem [{\citenamefont {Hahn}\ \emph {et~al.}(2005)\citenamefont {Hahn},
  \citenamefont {De~Bie}, \citenamefont {Stajich}, \citenamefont {Nguyen},\
  and\ \citenamefont {Cristianini}}]{hahn2005estimating}%
  \BibitemOpen
  \bibfield  {author} {\bibinfo {author} {\bibfnamefont {M.~W.}\ \bibnamefont
  {Hahn}}, \bibinfo {author} {\bibfnamefont {T.}~\bibnamefont {De~Bie}},
  \bibinfo {author} {\bibfnamefont {J.~E.}\ \bibnamefont {Stajich}}, \bibinfo
  {author} {\bibfnamefont {C.}~\bibnamefont {Nguyen}}, \ and\ \bibinfo {author}
  {\bibfnamefont {N.}~\bibnamefont {Cristianini}},\ }\href@noop {} {\bibfield
  {journal} {\bibinfo  {journal} {Genome Research}\ }\textbf {\bibinfo {volume}
  {15}},\ \bibinfo {pages} {1153} (\bibinfo {year} {2005})}\BibitemShut
  {NoStop}%
\bibitem [{\citenamefont {Gabetta}\ and\ \citenamefont
  {Regazzini}(2010)}]{gabetta2010gene}%
  \BibitemOpen
  \bibfield  {author} {\bibinfo {author} {\bibfnamefont {E.}~\bibnamefont
  {Gabetta}}\ and\ \bibinfo {author} {\bibfnamefont {E.}~\bibnamefont
  {Regazzini}},\ }\href@noop {} {\bibfield  {journal} {\bibinfo  {journal}
  {Mathematical Models and Methods in Applied Sciences}\ }\textbf {\bibinfo
  {volume} {20}},\ \bibinfo {pages} {1005} (\bibinfo {year}
  {2010})}\BibitemShut {NoStop}%
\bibitem [{sup({\natexlab{a}})}]{supp1}%
  \BibitemOpen
  \href@noop {} {} ({\natexlab{a}}),\ \bibinfo {note} {supplementary Material I
  Main Model Equations Derivation}\BibitemShut {NoStop}%
\bibitem [{sup({\natexlab{b}})}]{supp4}%
  \BibitemOpen
  \href@noop {} {} ({\natexlab{b}}),\ \bibinfo {note} {supplementary Material
  IV Supporting Material for Main Model}\BibitemShut {NoStop}%
\bibitem [{\citenamefont {Eigen}(2002)}]{eigen2002error}%
  \BibitemOpen
  \bibfield  {author} {\bibinfo {author} {\bibfnamefont {M.}~\bibnamefont
  {Eigen}},\ }\href@noop {} {\bibfield  {journal} {\bibinfo  {journal}
  {Proceedings of the National Academy of Sciences}\ }\textbf {\bibinfo
  {volume} {99}},\ \bibinfo {pages} {13374} (\bibinfo {year}
  {2002})}\BibitemShut {NoStop}%
\bibitem [{\citenamefont {Ewens}(1972)}]{ewens1972sampling}%
  \BibitemOpen
  \bibfield  {author} {\bibinfo {author} {\bibfnamefont {W.~J.}\ \bibnamefont
  {Ewens}},\ }\href@noop {} {\bibfield  {journal} {\bibinfo  {journal}
  {Theoretical Population Biology}\ }\textbf {\bibinfo {volume} {3}},\ \bibinfo
  {pages} {87} (\bibinfo {year} {1972})}\BibitemShut {NoStop}%
\bibitem [{sup({\natexlab{c}})}]{supp5}%
  \BibitemOpen
  \href@noop {} {} ({\natexlab{c}}),\ \bibinfo {note} {supplementary Material V
  Exponential Model}\BibitemShut {NoStop}%
\bibitem [{sup({\natexlab{d}})}]{supp2}%
  \BibitemOpen
  \href@noop {} {} ({\natexlab{d}}),\ \bibinfo {note} {supplementary Material
  II Spatial Epidemics Model}\BibitemShut {NoStop}%
\bibitem [{sup({\natexlab{e}})}]{supp3}%
  \BibitemOpen
  \href@noop {} {} ({\natexlab{e}}),\ \bibinfo {note} {supplementary Material
  III Network Model}\BibitemShut {NoStop}%
\bibitem [{\citenamefont {Mandelbrot}(1982)}]{Levyflights}%
  \BibitemOpen
  \bibfield  {author} {\bibinfo {author} {\bibfnamefont {B.~B.}\ \bibnamefont
  {Mandelbrot}},\ }\href@noop {} {\emph {\bibinfo {title} {The Fractal Geometry
  of Nature}}}\ (\bibinfo  {publisher} {New York: W. H. Freeman.},\ \bibinfo
  {year} {1982})\BibitemShut {NoStop}%
\end{thebibliography}
%



\section*{Supplementary material}

# Family-size variability grows with collapse rate in Birth-Death-Catastrophe model: Supplementary Material

N. Dori,[1] H. Behar,[2] H. Brot,[3] and Y. Louzoun[1, 4, *]

[1]*Gonda Brain Research Center, Bar-Ilan University, Ramat Gan, Israel*
[2]*Department of Biology, Stanford University Stanford, CA 94305-5020*
[3]*Boston Childrens Hospital, Harvard Medical School, 3 Blackfan Circle, Boston,
MA Center for Polymer Studies and Physics Department, Boston University, Boston, MA, 02215, USA*
[4]*Department of Mathematics, Bar-Ilan University, Ramat Gan, Israel*

(Dated: March 4, 2018)

This supplementary material includes five sections. In the first section we give detailed mathematical derivations of the original model equations. In the second and third sections we briefly introduce two additional models: the spatial epidemics model and the network model. These models support the results of the original model, namely, that catastrophes contribute to the system's inhomogeneity. In the forth section we give supplementary material regarding the original model. In the fifth and last section we give a graph that supports our claims regarding the the exponential model we refer to at the end of the main text.

## I. ORIGINAL MODEL

### A. Derivation of differential equations for $\frac{dN_k}{dt}$

Since the system is linear, it is possible to consider the contribution of each reaction separately. For convenience we write $\frac{dN_k}{dt} = \frac{dN_k^{b+m}}{dt} + \frac{dN_k^d}{dt} + \frac{dN_k^c}{dt}$, where $\frac{dN_k^{b+m}}{dt}$ refers to the contribution of birth+mutation processes, $\frac{dN_k^d}{dt}$ refers to the contribution of the death process and $\frac{dN_k^c}{dt}$ refers to the catastrophes contribution. The mutation is naturally integrated into the birth process, therefore birth and mutation processes are considered in the first term together.

#### 1. The birth and mutation contribution, $\frac{dN_k^{b+m}}{dt}$

When an individual is born, it must be added either to a new family or to an existing family. A birth into a new family creates a new family of size 1, and is, therefore, a mutation. This term appears only in the equation for $N_1$. A birth into an existing family could happen into any family size, $k \geq 1$, and it increases the family size by 1. Such birth events increase or decrease $N_k$ as follows:

1. If an individual is born to a family of size $(k-1)$, the family size increases to $k$ and $N_k$ increases, $N_k(t) = N_k(t-1) + 1$.

2. In parallel, if an individual is born to a family of size $k$, the family size increases to $(k+1)$ and $N_k(t) = N_k(t-1) - 1$.

In both cases the overall number of individuals is increased by 1, $m_1(t) = m_1(t-1) + 1$. The rate of a birth into a $k-1$ family is proportional to the product of three terms: the overall number of people in these families $((k-1)N_{k-1})$, the probability of an individual "giving birth" $\alpha$, and the probability of this birth not being a mutation $(1-\mu)$. It is more convenient, as will be apparent later, to multiply by $\frac{m_1}{m_1}$. Thus we have the first term in Eq. S1. The loss term in $N_k(t)$ is constructed in a similar way, and forms the second term in Eq. S1. It decreases $N_k$, thus the negative prefactor.

$$\frac{dN_k^{b+m}}{dt} = m_1\left[\alpha(1-\mu)(k-1)\frac{N_{k-1}}{m_1} - \alpha(1-\mu)k\frac{N_k}{m_1}\right] \quad \text{(S1)}$$

The above equation holds for $k > 1$. The $\frac{dN_k^{b+m}}{dt}$ for $k = 1$ is somewhat different. First, whenever a mutation birth occurs, $N_1$ increases by 1. This adds a positive term $\alpha\mu$. This is the rate of mutation births, and it is independent of $N_1$. Second, since $k = 0$ is not a family, the term responsible for births in a $k-1$ family does not exist here. We are left with Eq. S2

$$\frac{dN_1^{b+m}}{dt} = m_1\left[\alpha\mu - \alpha(1-\mu)\frac{N_1}{m_1}\right] \quad \text{(S2)}$$

#### 2. The death contribution, $\frac{dN_k^d}{dt}$

The death terms are constructed in a very similar way to the birth terms. When an individual dies, its family decreases by 1. We consider first families with $k > 1$. A death increases or decreases $N_k$ according to the following:

1. If an individual in a family of size $k$ dies, the family size decreases to $(k-1)$, and $N_k(t) = N_k(t-1) - 1$.

2. In parallel, if an individual in a family of size $(k+1)$ dies, the family size decreases to $k$, and $N_k(t) = N_k(t-1) + 1$.

---

* Corresponding author: louzouy@math.biu.ac.il

In both cases the overall number of individuals is decreased by 1, $m_1(t) = m_1(t-1) - 1$. Considering first families with $k > 1$, the rate of a death in a family of size $k$ is proportional to the overall number of people in these families, $(kN_k)$, multiplied by the probability of an individual dying, $\delta$. We again multiply by $\frac{m_1}{m_1}$. Thus we have the first term in Eq. S3. This term, as expected, tends to decrease $N_k$, thus the negative prefactor. The gain term in $N_k(t)$ is constructed in a similar way, and forms the second term in Eq. S3.

$$\frac{dN_k^d}{dt} = m_1\left[-\delta k \frac{N_k}{m_1} + \delta(k+1)\frac{N_{k+1}}{m_1}\right] \quad (S3)$$

We may replace $\delta$ by $\frac{m_1}{N}$ and get:

$$\frac{dN_k^d}{dt} = m_1\left[-k\frac{N_k}{\bar N} + (k+1)\frac{N_{k+1}}{\bar N}\right] \quad (S4)$$

The contribution of the death process to $\frac{dN_1}{dt}$ has exactly the same terms as for $k > 1$. Note that a death in $k = 1$ causes the family to disappear from the system, decreasing the total size of the system by 1, but requires no special treatment. Therefore we may arrive to the $\frac{dN_1}{dt}$ by substituting $k = 1$,

$$\frac{dN_1^d}{dt} = m_1\left[-\frac{N_1}{\bar N} + 2\frac{N_2}{\bar N}\right]. \quad (S5)$$

### 3. The catastrophes contribution, $\frac{dN_k^c}{dt}$

The rate of catastrophes to families of size $k$ is proportional to $\gamma$, the independent catastrophe rate, and to the frequency of such families in the system, $\frac{N_k}{m_0}$. It is multiplied by $m_1$ to describe the number of individuals killed in the catastrophe and added with a negative prefactor to indicate that it tends to decrease $N_k$.

$$\frac{dN_k^c}{dt} = m_1\left[-\gamma \frac{N_k}{m_0}\right] \quad (S6)$$

This term is applicable to $k \geq 1$, but for clarity, we write the $k = 1$ term explicitly,

$$\frac{dN_1^c}{dt} = m_1\left[-\gamma \frac{N_1}{m_0}\right] \quad (S7)$$

As explained in the main text, the total number of individuals, $m_1$, is recalculated and the death rate is set to $\delta = \frac{m_1}{N}$ at every time step. After $\delta$ is determined, $\alpha$, $\delta$ and $\gamma$ are normalized by the sum $\alpha + \delta + \gamma$, so that the sum of all effective rates is 1. This adds the prefactor $\frac{1}{\alpha+\delta+\gamma}$ to the master equations. To simplify our notation we do not write this prefactor in the master equation. It contributes only a time scaling to the results and does not affect the steady state.

### 4. Final equation $\frac{dN_k}{dt}$

Adding all these contributions together we arrive at the following equation.

$$\begin{cases} \dfrac{dN_1}{dt} = m_1\Big[\mu\alpha - \dfrac{\alpha(1-\mu)}{m_1}N_1 \\ \qquad + \dfrac{1}{\bar N}\big(-N_1 + 2N_2\big) \\ \qquad - \dfrac{\gamma N_1}{m_0}\Big] \\ \dfrac{dN_k}{dt} = m_1\Big[\dfrac{\alpha(1-\mu)}{m_1}\big((k-1)N_{k-1} \quad \text{for } k > 1 \\ \qquad - kN_k\big) \\ \qquad + \dfrac{1}{\bar N}\big(-kN_k + (k+1)N_{k+1}\big) \\ \qquad - \dfrac{\gamma N_k}{m_0}\Big] \end{cases} \quad (S8)$$

## B. Developing $\frac{dm_j}{dt}$

All moment differential equations begin from the definition,

$$m_j = \sum_k k^j N_k, \ j = 0, 1, 2. \quad (S9)$$

We differentiate with respect to time,

$$\frac{dm_j}{dt} = \sum_k k^j \frac{dN_k}{dt}, \ j = 0, 1, 2. \quad (S10)$$

and continue with inserting Eqs. S8.

### 1. Developing $\frac{dm_0}{dt}$

Inserting $j = 0$ in Eq. S10 we get

$$\frac{dm_0}{dt} = \sum_k k^0 \frac{dN_k}{dt} = \sum_k \frac{dN_k}{dt}. \quad (S11)$$

and

$$\begin{aligned}\frac{dm_0}{dt} =& m_1\Big[\mu\alpha - \alpha(1-\mu)\frac{N_1}{m_1} \\ & - \frac{N_1}{\bar N} + \frac{2N_2}{\bar N} - \frac{\gamma N_1}{m_0} \\ & + \sum_{k=2}^{k=\infty}\Big[\alpha(1-\mu)(k-1)\frac{N_{k-1}}{m_1} \\ & - \alpha(1-\mu)k\frac{N_k}{m_1} \\ & - \frac{kN_k}{\bar N} + \frac{(k+1)N_{k+1}}{\bar N} - \frac{\gamma N_k}{m_0}\Big]\Big]\end{aligned} \quad (S12)$$

All birth terms cancel when we perform the sum. Also most death terms cancel, except for a single term,

$$\frac{dm_0}{dt} = m_1\left[\mu\alpha - \frac{N_1}{\bar{N}} - \sum_{k=1}^{k=\infty}\frac{\gamma N_k}{m_0}\right] \tag{S13}$$

Using again the definition of $m_0$, Eq. S9, we get

$$\frac{dm_0}{dt} = m_1\left[\mu\alpha - \frac{N_1}{\bar{N}} - \gamma\right] \tag{S14}$$

### 2. Developing $\frac{dm_1}{dt}$

Inserting $j = 1$ in Eq. S10 we get

$$\frac{dm_1}{dt} = \sum_{k=1}^{\infty} k\frac{dN_k}{dt}. \tag{S15}$$

and

$$\begin{aligned}\frac{dm_1}{dt} =& m_1\Big[\mu\alpha - \alpha(1-\mu)\frac{N_1}{m_1} \\ &- \frac{N_1}{\bar{N}} + \frac{2N_2}{\bar{N}} - \frac{\gamma N_1}{m_0} \\ &+ \sum_{k=2}^{k=\infty} k\Big[\alpha(1-\mu)(k-1)\frac{N_{k-1}}{m_1} \\ &- \alpha(1-\mu)k\frac{N_k}{m_1} \\ &- \frac{kN_k}{\bar{N}} + \frac{(k+1)N_{k+1}}{\bar{N}} \\ &- \frac{\gamma N_k}{m_0}\Big]\Big]\end{aligned} \tag{S16}$$

This time the birth terms do not cancel completely, but rather

$$\begin{aligned}\frac{dm_1}{dt} =& m_1\Big[\mu\alpha + \frac{\alpha(1-\mu)}{m_1}\Big[-N_1 + \sum_{k=2}^{k=\infty}k(k-1)N_{k-1} \\ &- \sum_{k=2}^{k=\infty}k^2 N_k\Big] + \frac{1}{\bar{N}}\Big[\sum_{k=1}^{k=\infty}-k^2 N_k \\ &+ \sum_{k=1}^{k=\infty}(k+1)kN_{k+1}\Big] - \sum_{k=1}^{k=\infty}\frac{\gamma k N_k}{m_0}\Big].\end{aligned} \tag{S17}$$

Changing summation variable so that only $N_k$ appears in the equation,

$$\begin{aligned}\frac{dm_1}{dt} =& m_1\Big[\mu\alpha + \frac{\alpha(1-\mu)}{m_1}\Big[-N_1 + 2N_1 + \sum_{k=2}^{k=\infty}(k+1)kN_k \\ &- \sum_{k=2}^{k=\infty}k^2 N_k\Big] + \frac{1}{\bar{N}}\Big[\sum_{k=1}^{k=\infty}-k^2 N_k \\ &+ \sum_{k=2}^{k=\infty}k(k-1)N_k\Big] - \sum_{k=1}^{k=\infty}\frac{\gamma k N_k}{m_0}\Big].\end{aligned} \tag{S18}$$

When changing summation variables here and in the following, we neglect $N_k, k \to \infty$. Since the population is finite on average, $N_k \to 0$ exponentially as $k \to \infty$.

Next, higher order $k$ terms are cancelled,

$$\begin{aligned}\frac{dm_1}{dt} =& \mu\alpha + \frac{\alpha(1-\mu)}{m_1}\Big[N_1 + \sum_{k=2}^{k=\infty}kN_k\Big] \\ &+ \frac{1}{\bar{N}}\Big[-N_1 + \sum_{k=2}^{k=\infty}-kN_k\Big] \\ &- \sum_{k=1}^{k=\infty}\frac{\gamma k N_k}{m_0},\end{aligned} \tag{S19}$$

and substituting sums over $N_k$ by moments we get

$$\begin{aligned}\frac{dm_1}{dt} =& \mu\alpha + \frac{\alpha(1-\mu)}{m_1}\Big[m_1\Big] \\ &+ \frac{1}{\bar{N}}\Big[-m_1\Big] \\ &- \gamma\frac{m_1}{m_0},\end{aligned} \tag{S20}$$

And finally

$$\frac{dm_1}{dt} = m_1\left[\alpha - \frac{m_1}{\bar{N}} - \gamma\frac{m_1}{m_0}\right], \tag{S21}$$

### 3. Developing $\frac{dm_2}{dt}$

Inserting $j = 2$ in Eq. S10 we get

$$\frac{dm_2}{dt} = \sum_{k=1}^{\infty} k^2 \frac{dN_k}{dt}. \tag{S22}$$

and

$$\begin{aligned}\frac{dm_2}{dt} =& m_1\Big[\mu\alpha - \alpha(1-\mu)\frac{N_1}{m_1} - \frac{N_1}{\bar{N}} + \frac{2N_2}{\bar{N}} - \gamma\frac{N_1}{m_0} \\ &+ \sum_{k=2}^{\infty}k^2\Big[\alpha(1-\mu)(k-1)\frac{N_{k-1}}{m_1} - \alpha(1-\mu)k\frac{N_k}{m_1} \\ &- k\frac{N_k}{\bar{N}} + (k+1)\frac{N_{k+1}}{\bar{N}} - \gamma\frac{N_k}{m_0}\Big]\Big]\end{aligned} \tag{S23}$$

Whenever the $k = 1$ terms are identical to the higher $k$ terms they are inserted into the sum,

$$\frac{dm_2}{dt} = m_1\Big[\mu\alpha + \frac{\alpha(1-\mu)}{m_1}\Big[\sum_{k=2}^{\infty}k^2(k-1)N_{k-1} - \sum_{k=1}^{\infty}k^3 N_k\Big]$$
$$+ \frac{1}{N}\sum_{k=1}^{\infty}\Big[-k^3 N_k + k^2(k+1)N_{k+1}\Big]$$
$$- \gamma\sum_{k=1}^{\infty}k^2\frac{N_k}{m_0}\Big] \quad (S24)$$

A change of variable for some sums rearranges the terms so that we are left with $N_k$ only

$$\frac{dm_2}{dt} = m_1\Big[\mu\alpha + \frac{\alpha(1-\mu)}{m_1}\Big[\sum_{k=1}^{\infty}(k+1)^2 k N_k - \sum_{k=1}^{\infty}k^3 N_k\Big]$$
$$+ \frac{1}{N}\Big[\sum_{k=1}^{\infty}-k^3 N_k + \sum_{k=2}^{\infty}k(k-1)^2 N_k\Big]$$
$$- \gamma\frac{m_2}{m_0}\Big] \quad (S25)$$

$$\frac{dm_2}{dt} = m_1\Big[\mu\alpha + \frac{\alpha(1-\mu)}{m_1}\Big[\sum_{k=1}^{\infty}(k+1)^2 k N_k - \sum_{k=1}^{\infty}k^3 N_k\Big]$$
$$+ \frac{1}{N}\Big[-\sum_{k=1}^{\infty}k^3 N_k + \sum_{k=2}^{\infty}k(k-1)^2 N_k\Big]$$
$$- \gamma\frac{m_2}{m_0}\Big] \quad (S26)$$

$k^3$ birth terms are readily cancelled, while the death terms need one more step. However, since starting the summation from $k = 1$ instead of $k = 2$ adds a zero term to the sum, we can write,

$$\frac{dm_2}{dt} = m_1\Big[\mu\alpha + \frac{\alpha(1-\mu)}{m_1}\Big[\sum_{k=1}^{\infty}(2k^2 + k)N_k\Big]$$
$$+ \frac{1}{N}\Big[\sum_{k=1}^{\infty}-k^3 N_k + \sum_{k=1}^{\infty}k(k^2 - 2k + 1)N_k\Big]$$
$$- \gamma\frac{m_2}{m_0}\Big] \quad (S27)$$

$$\frac{dm_2}{dt} = m_1\Big[\mu\alpha + \frac{\alpha(1-\mu)}{m_1}\Big[\sum_{k=1}^{\infty}(2k^2 + k)N_k\Big]$$
$$+ \frac{1}{N}\Big[\sum_{k=1}^{\infty}(-2k^2 + k)N_k\Big]$$
$$- \gamma\frac{m_2}{m_0}\Big] \quad (S28)$$

$$\frac{dm_2}{dt} = m_1\Big[\mu\alpha + \frac{\alpha(1-\mu)}{m_1}(2m_2 + m_1)$$
$$+ \frac{1}{N}(m_1 - 2m_2) - \gamma\frac{m_2}{m_0}\Big] \quad (S29)$$

$$\frac{dm_2}{dt} = m_1\Big[\alpha + 2\alpha(1-\mu)\frac{m_2}{m_1}$$
$$+ \frac{1}{N}(m_1 - 2m_2) - \gamma\frac{m_2}{m_0}\Big] \quad (S30)$$

Finally, collecting all terms multiplied by $m_2$, we reach to the final equation

$$\frac{dm_2}{dt} = m_1\Big[\alpha + \frac{m_1}{\bar{N}} + m_2\Big[\frac{2\alpha(1-\mu)}{m_1}$$
$$- \frac{2}{\bar{N}} - \frac{\gamma}{m_0}\Big]\Big] \quad (S31)$$

## II. SPATIAL EPIDEMICS MODEL

In this section we briefly introduce a stochastic spatial model with births, competition, epidemic and diffusion events. The epidemics in this model, similar to the catastrophes in the original model, increase the inhomogeneity of the system.

This is a one-dimensional spatial model with $N$ sites. At each site $i$ there is one population, a family, with a discrete number of agents $B_i$. These families have a birth rate $\alpha$, a diffusion rate $D$, and a competition rate $\varepsilon$ (equal for all sites). $B_i$ can be described by the following equation:

$$\frac{dB_i}{dt} = \alpha B_i + D\Big(\frac{1}{2}(B_{i+1} + B_{i-1}) - B_i\Big) - \varepsilon B_i^2. \quad (S32)$$

In addition, at each time step there is a chance $\gamma$ that the entire family at site $i$ goes to extinction. $\gamma$ is called the epidemics rate. The diffusion is from site $i$ to its nearest neighbors $i - 1$ and $i + 1$ with periodic boundary conditions. All reactions are independent random events.

Monte Carlo simulations of this model were performed on a one-dimensional $1X1000$ lattice. We initiated the agents at random positions and enacted each reaction independently of the other reactions. At each site we computed the probability of each reaction and performed reactions according to the prescribed probabilities. In each time step all lattice sites were updated. The dynamics were simulated for different parameter values. The simulation framework was described in detail in previous publications [1, 2].

In figures S1 and S2 the simulations results for different values of $\gamma$ and $D$ are plotted. Figures S1 left and S2 left show the mean and variance of $B$ taken over different sites (different families) as a function of the diffusion rate

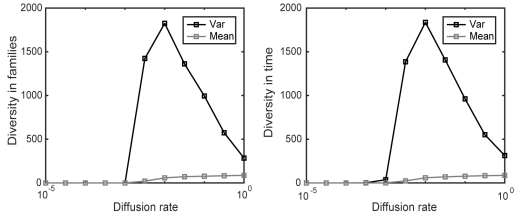

FIG. S1. Spatial epidemics model. Mean and variance of $B$, taken over families (left) and over time (right) as a function of the diffusion rate. The parameters that were used: $N = 1000$; $\alpha = 1; \varepsilon = 0.01; \gamma = 0.1$.

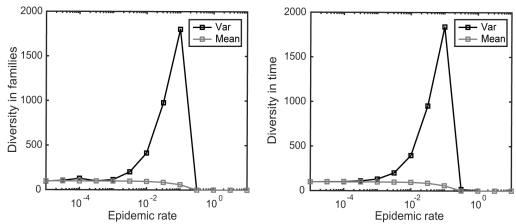

FIG. S2. Spatial epidemics model. Mean and variance of $B$, taken over families (left) and over time (right) as a function of epidemics rate $\gamma$. The parameters that were used: $N = 1000$; $\alpha = 1; \varepsilon = 0.01; D = 0.01$.

and the epidemics rate, respectively. Figures S1 right and S2 right show the mean and variance of $B$ taken over time as a function of the diffusion rate and the epidemic rate, respectively. The diffusion in the spatial epidemics model ($D$) corresponds to the mutation rate of the original model ($\mu$). The epidemic rate in the spatial epidemics model ($\gamma$) corresponds to the catastrophe rate of the original model ($\gamma$). In Fig. S1, for very large diffusion rates the diversity of the population is low. Decreasing the diffusion rate increases the diversity of the population. For very small diffusion rates the entire population goes to extinction. These dynamics are similar to the one in the original model, if the diffusion rate is replaced by the mutation rate (see Fig. (1) in the main text). In Fig. S2, for very low epidemic rates, the diversity of the population is low. Increasing the epidemics rate increases the diversity. For a very high epidemics rate the entire population goes to extinction (the diversity goes to zero). These results are similar to the results of original model, most notably, that increasing the catastrophe/epidemics rate increases rather than decreases system diversity.

## III. NETWORK MODEL

In this section we briefly introduce a network generation model where we compare two types of edge deletion:

1. Deletion of a single a edge
2. Deletion of all edges connected to a node.

Where the overall number of edges is constant. An edge deletion is comparable to a single death in the original model, while deletion of all edges belonging to a node is comparable to a catastrophe in the original model. In the first case the edge to be deleted is chosen randomly, in the second case, the node whose edges are to be deleted is chosen randomly. The total number of edges is kept constant by adding the same number of edges as has been deleted.

We check two types of edge addition.

(A) Linear preferential attachment - A node is chosen with probability $c + 1$, where $c$ is the node degree. A second node is chosen randomly, and a new edge is created between the pair if they are unconnected. If the pair is already connected the processes is started afresh until an unconnected pair is chosen and a new edge is added between them.

(B) Triadic closure - A node is chosen randomly and an edge is added between two of its neighbors. The algorithm is as follows. A node is chosen randomly. Then, two of its neighbors are chosen randomly. If an edge already exists between these two neighbors, we start afresh from the beginning. If the first node has less then 2 neighbors, a new edge is created between that node and another randomly selected node.

We compare single edge deletion and multiple edges deletion in both types of edge addition. See [3] for model and simulation details.

The degree distributions of a random Erdos-Renyi networks in equilibrium are given in Fig. S3. Upper and lower plots correspond to edge addition types (A) and (B) respectively. In both types of edge addition, the degree distribution of deletion of all edges of a node (catastrophes) has a fatter tail compared with a single edge deletion, as was the original model.

## IV. SUPPORTING MATERIAL FOR ORIGINAL MODEL

The data in this section refers to the original model (main text). In Fig. S4 we compare simulated and theoretic $N_1$ values, and show that the simulation is in excellent agreement with theory. In Fig. S5, the time development of the moments $m_0$, $m_1$, and $m_2$ with two $\gamma$ values are compared. The higher variance of the system in the "high variance" phase is clear. In Table S1 we give the parameters details of the points on the black line of Fig. 2 (main text).

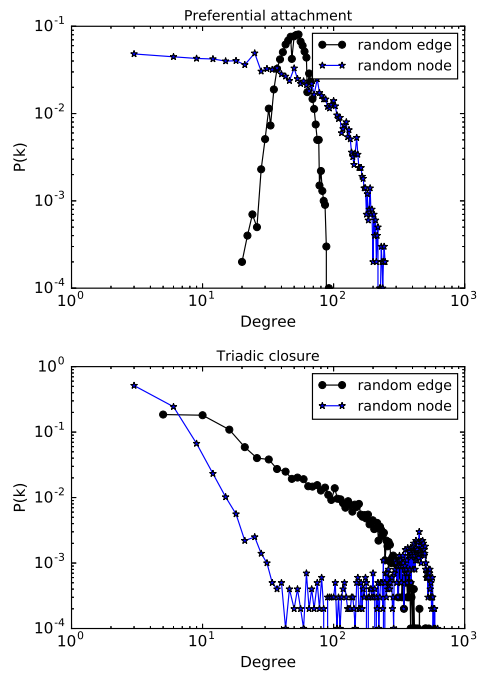

FIG. S3. Network model. Degree distribution of the two types of edge addition with either random edge deletion (black line) or the removal of all edges from a node (blue line). Upper plot - linear preferential attachment. Lower plot - triadic closure. As was the case in all previous systems studied, the simultaneous removal of all edges from a node increase the variance in the degree distribution (i.e. it leads to a fatter tail). In this case, we do not have a continuous variable for the removal, but rather a binary decision - only catastrophes or only single deaths.

## V. EXPONENTIAL MODEL

In this section we give the family size distribution of the exponential model S6. This family size distribution has a similar relation between $\gamma$ and the family size distribution variance as the original model, and it shows that system with higher catastrophe frequency also have higher frequency of large families.

[1] H. Behar, N. Shnerb, and Y. Louzoun, Physical Review E **86**, 031146 (2012).
[2] H. Behar, A. Agranovich, and Y. Louzoun, Mathematical biosciences and engineering: MBE **10**, 523 (2013).
[3] H. Brot, M. Honig, L. Muchnik, J. Goldenberg, and Y. Louzoun, Physical Review E **88**, 042815 (2013).

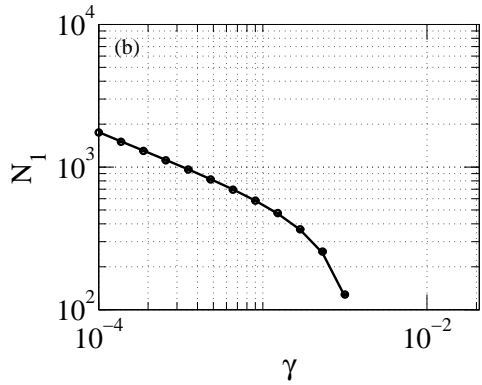

FIG. S4. Comparison of simulation and theory, original model. $N_1$ is the number of families of size 1. Each point represents a dot on the black line of Fig. 2. See Table S1 for parameters. Empty regions represent values of 0. Each simulation result is the average over 5 realizations and 50 time points separated by $10^6$ time steps after steady-state has been established. The lines are the theory and the dots are simulation. Theory and simulation are in perfect fit.

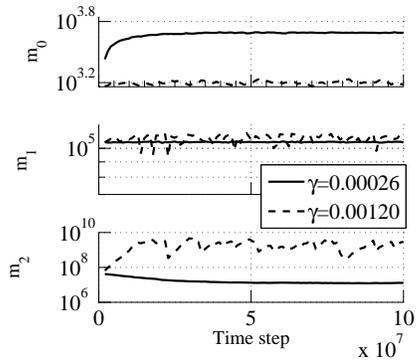

FIG. S5. Dependence of simulated $m_0$, $m_1$, and $m_2$ on time. The first system $\gamma = 0.00026$ (line 4 of Table S1) is in the low diversity phase. The second system $\gamma = 0.00120$ (line 9 of Table S1) is in the high diversity phase. Results are averaged over 5 replicas. The high diversity of the second system is evident in $m_0$, $m_1$, and $m_2$.

| | $\gamma$ | $\mu$ | $\alpha$ | $N$ | Phase | $\eta$ | $k_{\max}$ |
|---|---|---|---|---|---|---|---|
| 1 | 0.00010 | 0.01700 | 1.0099 | 102860 | $B<0$ | 1.0290 | 66 |
| 2 | 0.00014 | 0.01465 | 1.0136 | 102860 | $B<0$ | 1.0285 | 77 |
| 3 | 0.00020 | 0.01260 | 1.0186 | 102870 | $B<0$ | 1.0305 | 90 |
| 4 | 0.00026 | 0.01089 | 1.0254 | 102880 | $B<0$ | 1.0347 | 107 |
| 5 | 0.00040 | 0.00940 | 1.0348 | 102890 | $B<0$ | 1.0478 | 133 |
| 6 | 0.00048 | 0.00809 | 1.0476 | 102900 | $B<0$ | 1.0603 | 167 |
| 7 | 0.00070 | 0.00700 | 1.0651 | 102920 | $B>0$ | 1.0925 | 233 |
| 8 | 0.00090 | 0.00601 | 1.0892 | 102940 | $B>0$ | 1.1321 | 352 |
| 9 | 0.00120 | 0.00520 | 1.1221 | 102980 | $B>0$ | 1.1924 | 636 |
| 10 | 0.00169 | 0.00447 | 1.1671 | 103020 | $B>0$ | 1.2808 | 1664 |
| 11 | 0.00230 | 0.00390 | 1.2287 | 103090 | $B>0$ | 1.3653 | 5348 |
| 12 | 0.00316 | 0.00332 | 1.3131 | 103180 | $B>0$ | 1.3847 | 11587 |
| 13-22 | $\gamma > 0.0032$ on line | $\mu < 0.0033$ on line | $\alpha > 1.31$ | $N > 103180$ | $N_1 < 0$ | - | - |

TABLE S1. Parameters of dots on the black line of Fig. 2. The equation of the black line is $\log_{10}(\mu) = -0.4729\log_{10}(\gamma) - 3.6611$. Note that any other combination could have been chosen with similar results, as long as both phase transitions are traversed. $\eta$ and $k_{\max}$ are a result of the theoretic calculation.

| Parameter | Name |
|---|---|
| $\alpha$ | Birth rate |
| $\delta$ | Death rate |
| $\mu$ | Mutation rate |
| $\gamma$ | Catastrophe rate |
| $T$ | Total number of time steps, $10^8$ |

TABLE S2. Symbols definitions, original model

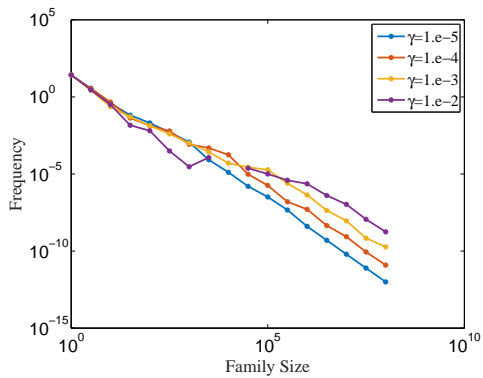

FIG. S6. Family-size distribution of the exponential model. The tail size increases as the catastrophe rate $\gamma$ increases. Here the bins are logarithmic, and $\alpha = 1$, $\mu = 1e-8$, $N = 1e8$.



\end{document}